\documentclass[english,showpacs,preprint,superscriptaddress]{revtex4-1}
\usepackage{lmodern}
\usepackage{lmodern}
\usepackage[T1]{fontenc}
\usepackage{amsmath}
\usepackage{amssymb}
\usepackage{graphicx}
\usepackage{esint}
\usepackage{ mathrsfs }
\usepackage{mhchem}
\makeatletter

\usepackage{amsthm}\usepackage{latexsym}\usepackage{bm}\usepackage{amsfonts}

\usepackage{babel}

\usepackage{babel}

\usepackage{hyperref}
\hypersetup{
	breaklinks = true,
    colorlinks = true,
    citecolor = {blue},
	urlcolor = {blue},
	linkcolor = {blue}
}

 \global\long\def\rhostar{ {\rho^\star}}

\global\long\def\EXP{\times10^}
\usepackage{cleveref}

 \global\long\def\GammaSP{{\Gamma}_\mathrm{SP}}

 \global\long\def\rmd{\mathrm{d}}

 \global\long\def\bfx{\mathbf{x}}
 
 \global\long\def\bfv{\mathbf{v}}

 \global\long\def\bfA{\mathbf{A}}
 
 \global\long\def\bfB{\mathbf{B}}
 \global\long\def\bfb{\mathbf{b}}

 \global\long\def\bfe{\mathbf{e}}

 \global\long\def\FIG#1{Fig.~\ref{#1}}
 \global\long\def\EQ#1{Eq.~(\ref{#1})}

 \global\long\def\REF#1{Ref.~\citep{#1}}

 \global\long\def\calD{\mathcal{D}}

\usepackage{babel}

\usepackage{babel}

\@ifundefined{showcaptionsetup}{}{%
 \PassOptionsToPackage{caption=false}{subfig}}
\usepackage{subfig}
\makeatother

\usepackage{babel}
\begin{document}

\title{Bohm-like Neoclassical Transport in Highly Collisional Toroidal Plasmas with High Density Gradients}

\author{Jianyuan Xiao}

\author{Huishan Cai}
\author{Jian Liu}
\email[Corresponding author: ]{jliuphy@ustc.edu.cn}

\affiliation{School of Nuclear Sciences and Technologys, University of Science and Technology
of China, Hefei, 230026, China}

\author{Zhi Yu}

\author{Yifeng Zheng}
\affiliation{Institute of Plasma Physics Chinese Academy of Sciences Hefei,230031,China}
\begin{abstract}
Conventional neoclassical theory in the 
Pfirsch-Schl\"{u}ter regime fails to 
accurately model collision-induced transport in toroidal plasmas with
high density gradients. In this scenario, we find that collision suppresses the
return flow, leading to the dominance of the transport flux by the vacuum 
toroidal field drift with a reduced Bohm-like scaling. The new 
regime is also confirmed by full-orbit particle simulations, and can be 
employed to improve the accurate modeling of impurity transport in 
toroidal magnetized plasmas.

\end{abstract}
\maketitle

Neoclassical transport \cite{PhysRevLett.29.698,doi:10.1063/1.1693728,hinton1976theory,helander2005collisional} is a fundamental 
theory used to calculate the transport of particles, momentum, 
and heat resulting from collisions in confined plasmas 
with complex magnetic geometries. It is usually applied in researches of transport in toroid
shaped magnetic confined plasmas such as Tokamak \cite{Hirshman_1981,hinton1976theory}, 
Reversed field 
pinch \cite{gobbin2010neoclassical}, Stellarator \cite{galeev1969plasma,beidler2003neoclassical} 
and even 
some space plasmas \cite{cunningham2018neoclassical}. In the field of tokamak
physics, neoclassical transport theory primarily focuses on modeling 
the behavior of fast particles and impurities \cite{Hirshman_1981,Goloborod_ko_1983,Angioni_2014,bader_drevlak_anderson_faber_hegna_likin_schmitt_talmadge_2019,Angioni_2021,kiptily2023evidence}.
In other magnetic confinement fusion devices like Stellarators,
understanding the neoclassical transport of the main particle species becomes crucial \cite{doi:10.1063/1.860843,doi:10.1063/1.3553025}.
One notable observation is that neoclassical diffusion coefficients are approximately an order of magnitude larger than their classical counterparts. One of the most significant and widely accepted results of neoclassical transport theory, derived around 50 years ago, is the identification of three distinct regimes (Pfirsch-Schl\"{u}ter, Plateau and Banana) \cite{hinton1976theory,helander2005collisional,beidler2003neoclassical,gobbin2010neoclassical} for the diffusion coefficient $D_\mathrm{NC}$ in toroidal magnetic confinement plasmas. These regimes are distinguished by relationship 
between the collisional frequency $\nu$, the inverse local aspect ratio $\epsilon$,
and the transit 
frequency $\omega_t=\frac{v_T}{qR_0}$, given the axisymmetric toroidal geometry with coordinate 
$\left( r,\theta,\phi \right)$ as
\begin{equation}
	\calD_\mathrm{NC}= \left\{
	\begin{array}{ccll}
		\calD_\mathrm{Banana}&\approx &2\epsilon^{-3/2}q^2\rho^2\nu,&\text{if } \nu \leq \omega_t\epsilon^{3/2}\\
		\calD_\mathrm{Plateau}&\approx &2q^2\rho^2\omega_t,&\text{if }\epsilon^{3/2}\omega_t<\nu \leq \omega_t\\
		\calD_\mathrm{PS}&\approx &2q^2\rho^2\nu,&\text{if } \omega_t\leq\nu
	\end{array}
	\right.. \label{EqnPPB}
\end{equation}
Here $v_T$ represents the thermal speed $\sqrt{T_i/m}$, where $T_i$ is the ion temperature 
and $m$ is the mass of the charged particle, $q$ denotes 
the safety factor, $R_0$ denotes
the major radius of 
the toroidal plasma, and $\epsilon=r/R_0$, $\rho=m_iv_T/(q_iB_0)$ represents the ion
gyro-radius at the thermal speed, $q_i$ is the charge of the ion.
These results are obtained through the solution of drift-kinetic approximation of 
Vlasov-Fokker-Planck (VFP) equations. Alternatively, they can be explained by analyzing the step-size of the drift-induced random walk and the effective collision frequency of the guiding center. 
However, we have discovered that in the Pfirsch-Schl\"{u}ter (PS) regime, there is a saturation 
of particle flux strength when the parameter
$G_\nu=2\nu R_0\kappa q^2 /\Omega_i$ approaches or exceeds 1. Here $\Omega_i$ is the ion
cyclotron frequency, $\kappa$ 
is the inverse of scale length of distribution function. 
This saturation phenomenon cannot be properly described by the neoclassical 
theory in the PS regime. As far as our knowledge extends, there is
currently no theoretical or numerical investigation addressing this issue at the kinetic level.

In this letter, we introduce a novel model for determining the collision-induced 
particle transport in toroidal plasmas with high density gradients and
high collisional frequencies.
In this regime, the conventional neoclassical transport theory 
for the Pfirsch-Schl\"{u}ter (PS)
regime cannot be applied, and we refer to it as the Second Plateau (SP) regime. 
The basic model is solving the VFP equation numerically or analytically 
in a ring domain of the poloidal cross section. Assuming that the 
width of ring is $\Delta r=1/\kappa$, 
the relation between particle flux, normalized gyro-radius
$\rho^\star=q\rho_T/\Delta r$ and $G_\nu\sim 1$ 
is calculated from the static solution under the drift-kinetic approximation of 
VFP equation in toroidal geometry obtained by the Lattice Boltzman Method (LBM). 
When $G_\nu \gg 1$, the particle flux is calculated from an approximate 
analytical solution of VFP.
To validate
the accuracy of the new model, we conduct 6D full-orbit particle simulations 
in a simplified tokamak field configuration. All results provide two clear conclusions:
1) As the threshold parameter $G_\nu$ approaches or 
exceeds $1$, the neoclassical transport theory fails to predict the particle flux in the PS regime. 
2) Under such circumstances,
the collision disrupts the parallel return flow (PS current) of guiding centers. 
According to approximate solution of the VFP equation in a ring domain of the
poloidal cross section with $G_\nu\gg 1$,
the flux across the magnetic field surface 
should comply 
\begin{eqnarray}
	\GammaSP=\frac{T_i}{q_iB_0R_0}\cdot\frac{4}{\pi}n_0\,,
\end{eqnarray}
where $B_0$ is the strength of toroidal magnetic field at the magnetic 
axis, $n_0$ is the density at $r=r_i$,
$\kappa=1/(r_o-r_i)$, $r_o$ and $r_i$ are 
radii of the outter and inner circle of the ring, respectively.
In this case, the flux is proportional to the guiding center flux 
generated by vacuum toroidal field (VTF) drift.
The findings of this study differ significantly from the predictions of conventional 
neoclassical theory in several ways:
1) The particle flux, induced by collisions, is hardly affected by
the inverse density gradient scale $\kappa$, collision 
frequency $\nu$, or safety factor $q$. 
This phenomenon can be interpreted as a type of transport barrier 
that arises in highly collisional toroidal plasmas with high density gradients.
2) The transport is solely driven by collisions and the magnetic field geometry; it occurs even in the absence of an electric field.
3) The resulting guiding center flux is 
proportional to $T_i/(B_0R_0)$, exhibiting a reduced Bohm-like scaling. A tokamak with a larger magnetic field or major radius can mitigate this type of transport.
It is worth mentioning that previous work by Helander et al. indicates that in the presence 
of high-density gradients for highly charged impurity ions, the transport of high-$Z$ 
(atomic number) impurity ions may not conform to neoclassical theory. In fact, 
the particle flux may even decrease with the increase of gradients 
\cite{helander1998bifurcated,fulop1999nonlinear,fulop2001nonlinear}. 
However, that theory is based on reduced fluid equations and requires the rotation of impurities, 
which distinguishes it from the present model.

\begin{figure}[htp]
	\begin{center}
		\includegraphics[width=0.6\linewidth]{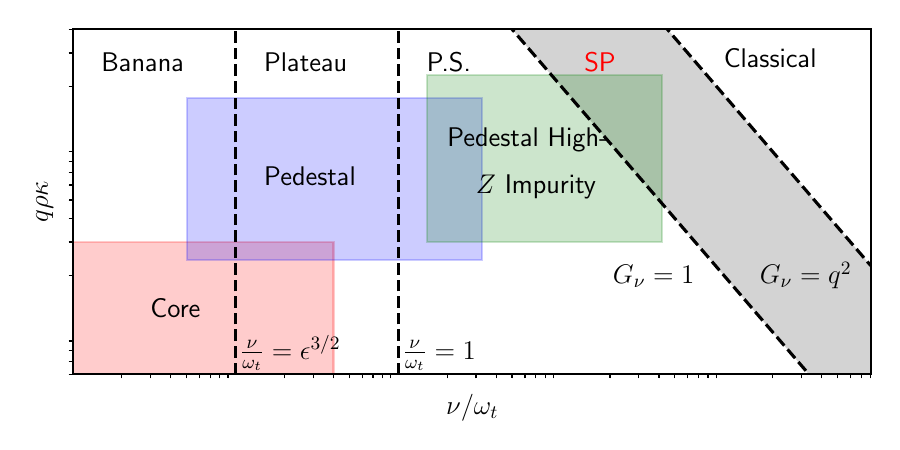}
	\end{center}
	\caption{Collision induced transport theories and their applicable regime in modern tokamak plasmas. The present SP model is valid for $1< G_\nu < q^2$ and $\nu\gg\omega_t$, which is presented in the gray area and high-$Z$ impurities in the pedestal of the plasma can satisfy this condition.}
	\label{FigParaNC}
\end{figure}

In modern tokamak plasmas, the ion temperature and background density generally fall within the range of $10^2\sim10^4\mathrm{eV}$ and $1\EXP{18}\sim1\EXP{20} \mathrm{m}^{-3}$, respectively.
The applicable collision-induced transport theories and corresponding regimes 
are illustrated in
\FIG{FigParaNC}. In addition to the conventional PS regime, the 
present model also applies to the Second Plateau (SP) regime, 
specifically when the parameter $G_\nu$ satisfies the condition $1<G_\nu<q^2$.
The present theory can be employed to explain the asymmetric distribution of impurities 
observed in some experiments \cite{terry1977observation,rice1997x,reinke2013non} of high density 
tokamak plasmas. It can also be used to more accurately model the transport of high-$Z$
impurities in toroidal magnetized plasmas, which is crucial since these impurities 
significantly affect the behavior of plasmas. Future high performance magnetic confinement
fusion reactors are designed based on these modelings.

Collision-induced particle transport can be obtained from the 
static solution of drift-kinetic approximation of the Vlasov-Fokker-Planck (VFP) 
equation \cite{hinton1976theory,lin1995gyrokinetic,helander2005collisional},
\begin{eqnarray}
	\dot{f}+\left( u\bfb+\bfv_\rmd \right)\cdot\nabla f-C\left( f \right)&=& 0\,,\label{EqnGCVFP}
\end{eqnarray}
where $f=f\left( \bfx,v,\xi,t \right)$ is the guiding center distribution 
function, $\bfx=\left( r,\theta,\phi \right)$ and $\bfv$
are the location and velocity of the guiding center, respectively,
$v=|\bfv|$, $\xi=\bfv\cdot\bfb/v$, $\bfB$ 
is the magnetic field, $B=|\bfB|$, $\bfb=\bfB/B$, $\bfv_\rmd$ is the drift velocity
\begin{eqnarray}
	\bfv_\rmd=\bfe_z\frac{v^2\left( 1+\xi^2 \right)}{2\Omega_iR_0}\,,
\end{eqnarray}
and $C$ is the drift kinetic collision operator. 
We consider the 
solution of \EQ{EqnGCVFP} in the following domain
\begin{eqnarray}
	r_i\leq r \leq r_o,\quad0\leq \theta \leq 2\pi\,,
\end{eqnarray}
with boundary condition as 
\begin{eqnarray}
	\int_0^1 f|_{r=r_i} \rmd \xi=f_i,\quad f|_{r=r_o}=0,\quad f|_{\theta=0}=f|_{\theta=2\pi}\,. \label{EqnBCGC}
\end{eqnarray}
This setup is a simplified model for the following situation. 
if $r\leq r_i$, the collisional frequency is low and 
particles are nearly uniformly distributed on the magnetic surface. If
$r\geq r_o$, the temperature is low and plasmas will become neutral gas, so charged
particles vanish at $r=r_o$. Though this setup is relatively rough, it is enough to 
demonstrate how charged particles transport in the regime where the collisional frequency 
and density gradient are both very high. 

\begin{figure}[htp]
	\begin{center}
		\subfloat[]{\includegraphics[width=0.53\linewidth]{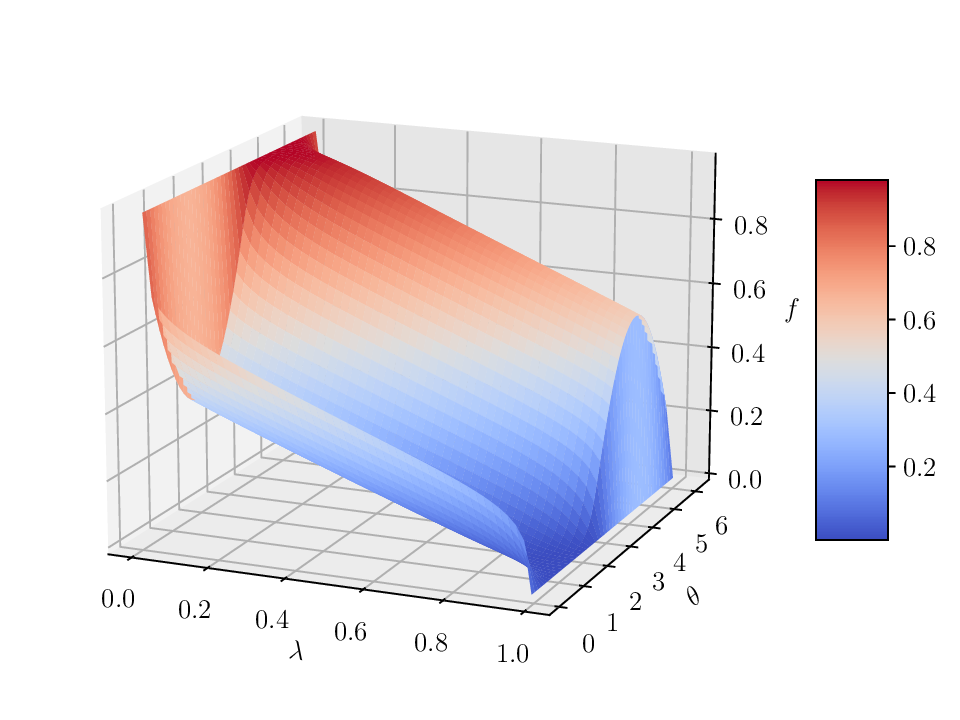}\label{FigFLUXNUMF3D}}
		\subfloat[]{\includegraphics[width=0.46\linewidth]{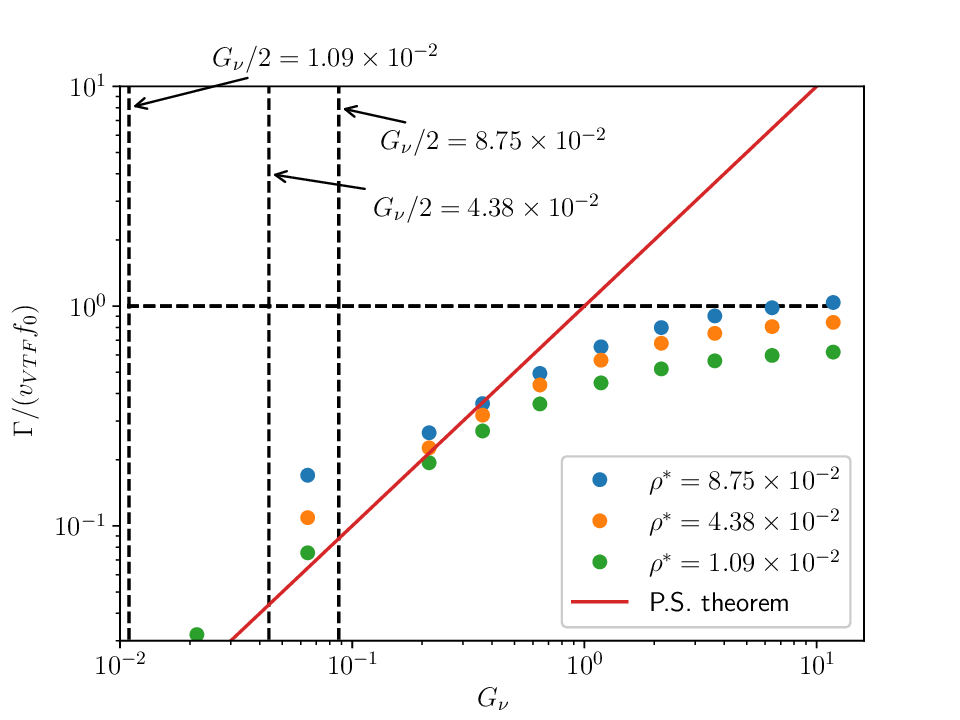}\label{FigFLUXNUMFLUX}}
	\end{center}
	\caption{The static solution of the VFP equation when $G_\nu=2$ (a) and the flux averaged particle flux versus $G_\nu$ obtained by the LBM method for different $G_\nu$ and $\rho^\star$ (b).}
	\label{FigFLUXNUM}
\end{figure}

We solve \EQ{EqnGCVFP} under boundary condition \EQ{EqnBCGC} numerically
using the 1st-order Lattice Boltzman Method (LBM). 
The collisional operator $C$ is choosen as the pitch
angle collisional operator
\begin{eqnarray}
	C\left( f \right)=\nu C_0\left( f \right)=\nu \frac{\partial}{\partial \xi}\left( \left( 1-\xi^2 \right)\frac{\partial f}{\partial \xi}  \right)\,.
\end{eqnarray}
The simulation stops when the distribution
function $f$ becomes stable, to reach a static solution. An example of such solution
$f\left( \lambda, \theta \right) = \int \rmd \xi f\left( \lambda,\theta,\xi \right)$ is shown in \FIG{FigFLUXNUMF3D}, where $G_\nu=2$ and $\lambda=\left( r-r_i \right)/\left( r_o-r_i \right)$.
Then we can calculate
the flux surface averaged particle flux through 
\cite{hinton1976theory,lin1995gyrokinetic,helander2005collisional}
\begin{eqnarray}
	\Gamma&=& \frac{\int_0^{2\pi}\rmd\theta h\int \rmd\xi \rmd v 4\pi v^2 f v_r}{\int_0^{2\pi}\rmd\theta h}~,\label{EqnGammaNUM}
\end{eqnarray}
where $h=1+\frac{r}{R_0}\cos\left( \theta \right)$.
The dependence of $\Gamma$ on $G_\nu$ and $\rhostar=\frac{qv}{\Omega_i\left( r_o-r_i \right)}$ are plotted in \FIG{FigFLUXNUMFLUX}.
The particle flux is smaller than the prediction of conventional neoclassical 
theory in the PS regime, and it saturates to a value in the same order as 
$f_0v_\mathrm{VTF}$ when $G_\nu \gg 1$. More detailed analysis shows that (see
the supplemental material) the flux surface averaged particle flux should be 
approximately
\begin{eqnarray}
	\GammaSP=\frac{v_\mathrm{th}^2}{\Omega_iR_0}\frac{4}{\pi}n_0\left( r_i \right)=\frac{T_i}{q_iB_0R_0}\frac{4}{\pi}n_0\left( r_i\right)~,\label{EqnANAFLUX}
\end{eqnarray}
when $G_\nu\rightarrow\infty$.

The neoclassical transport theory in the PS regime
\cite{hinton1976theory,lin1995gyrokinetic, helander2005collisional}
states that the diffusive coefficient is approximately $q^2$ times the classical one.
However, when $\kappa\sim1/\left( r_o-r_i \right)$ is so high that $G_\nu$ approaches 
or exceeds 1, the perturbed distribution function 
in the conventional neoclassical theorem for PS regime
$f_{PS,1}=-G_\nu n_0(r_i)f_M \sin\left( \theta \right)$ \cite{hinton1976theory,lin1995gyrokinetic,helander2005collisional} becomes comparable or even larger than
$n_0(r_i)f_M$, violating the assumption that $f_{PS,1}\ll n_0\left( r_i \right)f_M$. A similar condition where the neoclassical theory is not applicable is also 
mentioned in \REF{helander2005collisional}. 
In these cases, our result indicates that the particle flux will saturate to $\GammaSP$ as $G_\nu\rightarrow\infty$.
It is independent of the density gradient $n_0'$, 
collision frequency $\nu$, and the safety factor $q$. $\GammaSP=\frac{4 T_i}{\pi q_iB_0R_0}n_0$ 
indicates
that this particle flux applies in the Second Plateau (SP) regime 
independent of the collision frequency.
We recall the particle flux caused by Bohm diffusion $\Gamma_\mathrm{Bohm}$, which is
\begin{eqnarray}
	\Gamma_\mathrm{Bohm}\approx \frac{1}{16}\frac{T_i}{q_iB_0}n_0'\left( r_i \right)\,.
\end{eqnarray}
It is clear that both the SP particle flux and the Bohm diffusion particle 
flux are independent of the collision frequency 
$\nu$ and the safety factor $q$. And both of them are proportional
to $T_i/(q_iB_0)$. 
The only distinction is that the density gradient $n_0'$ in the Bohm diffusion
is replaced by $n_0/R_0$ in the SP model.

The SP regime is likely to occur at the pedestal of a tokamak plasma where 
the collisional frequency is high,
the density profile is steep 
and the safety factor $q$ is high. For example, in a tokamak pedestal 
plasma with
$B_0=2.5\mathrm{T}$, $R_0\kappa\sim100$, $q=3.5$, 
$T_i=200\mathrm{eV}$, $n_e=n_i=1\EXP{19}\mathrm{m}^{-3}$, $\log\left( \Lambda \right)\sim 20$, 
the major ion species is deterium,
we consider impurities such as $\mathrm{C}^{6+}$.
Due to an extra $Z^2$ factor in the collisional frequency \cite{Angioni_2021}, 
where $Z$ represents the charge state of the ion, we can estimate that
$G_{\nu_\mathrm{C}}\sim6^2G_{\nu_\mathrm{D}}\sim1.8$, which is bigger than 1.

\begin{figure}[htp]
	\begin{center}
		\subfloat[]{\includegraphics[width=0.49\linewidth]{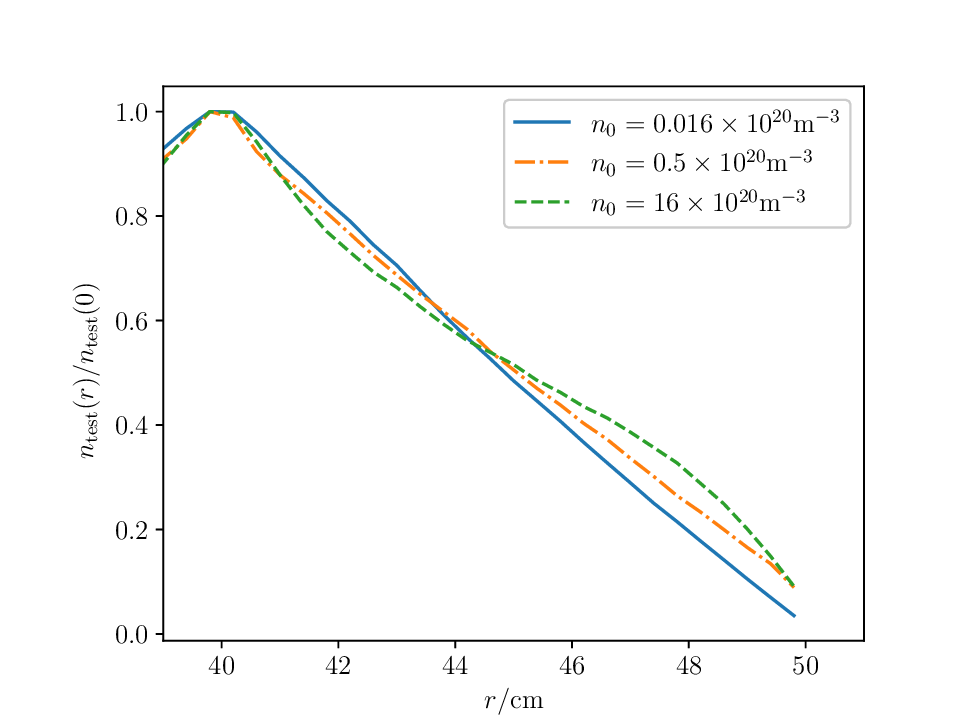}}
		\subfloat[]{\includegraphics[width=0.49\linewidth]{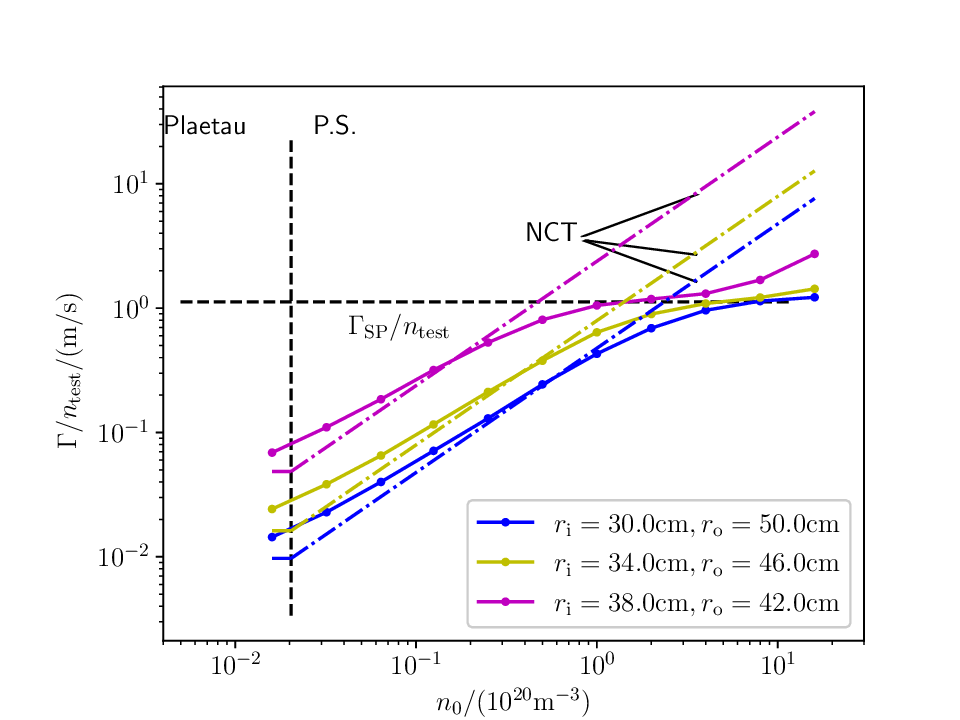}}\\
		\subfloat[]{\includegraphics[width=0.49\linewidth]{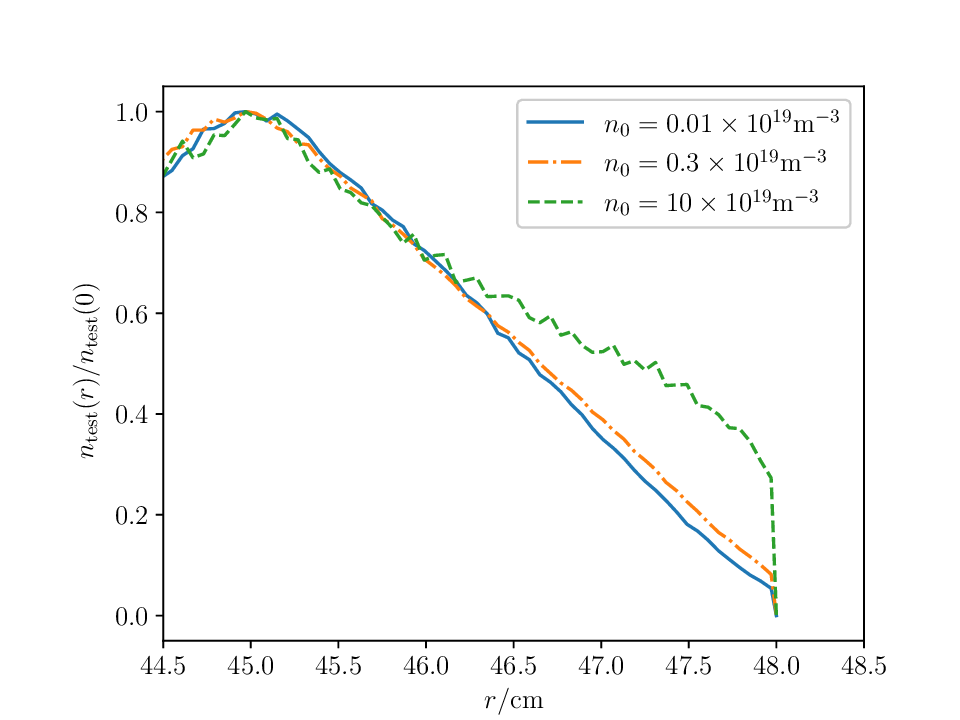}}
		\subfloat[]{\includegraphics[width=0.49\linewidth]{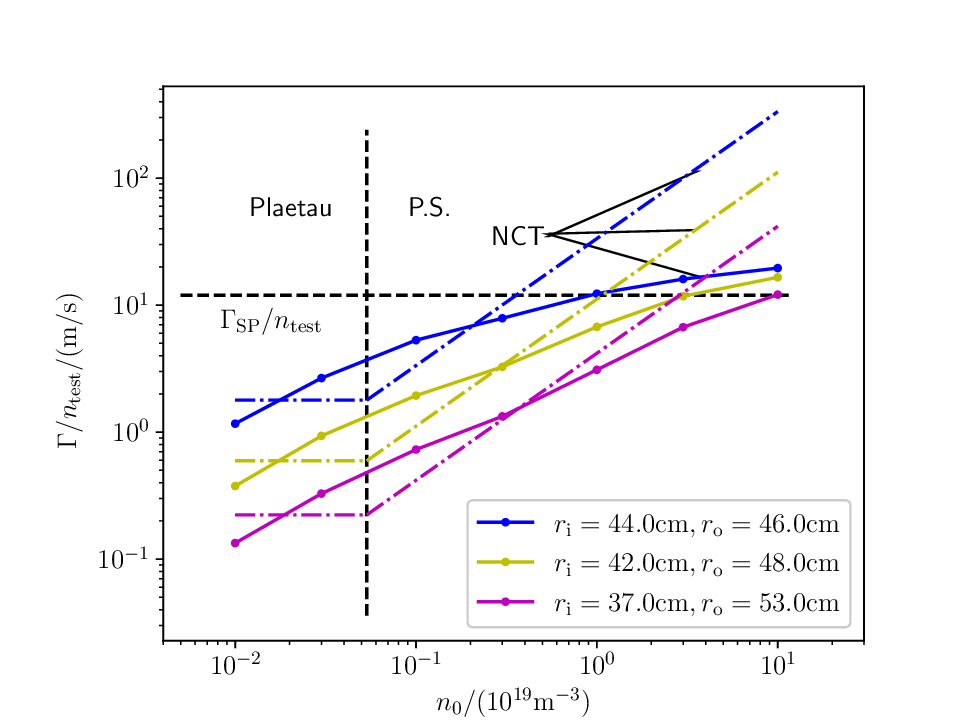}}
	\end{center}
	\caption{Static density profile for $r_o=50\mathrm{cm}$ in Case A (a), $r_o=48\mathrm{cm}$ in Case B (c), and particle flux for different backgroud densities and sink locations for case A (b) and B (d), where NCT means the neoclassical theorem and $\GammaSP$ is the new SP flux displayed in \EQ{EqnANAFLUX}. For $G_\nu>q^2$, the classical collisional effect will become significant and hence the particle flux increases with the collisional frequency again.}
	\label{FigFLUXVerifySTAT}
\end{figure}

Next, we validate the SP particle flux $\GammaSP$
for plasmas with high collisional frequency through 6D full-orbit 
particle tracing simulations. The particle evolution scheme 
employs the standard Boris method, while the Stratonovich stochastic differential 
equations for Coulomb collisions with background Maxwellian distributed ions 
are solved using the implicit midpoint method
\cite{zheng2021issde}.
The magnetic field
configuration chosen is the same as the one used in \REF{qin2009variational},
i.e., the vector
 and scalar potentials $\bfA$ and $\psi$ are
\begin{eqnarray*}
	\bfA\left(x,y,z\right)  =  \frac{1}{2}B_{0}\left(\frac{r^{2}}{qR}\bfe_{\phi}-\log(\frac{R}{R_0})R_0\bfe_{z}+\frac{R_0z}{R}\bfe_{R}\right)\,, \psi\left(x,y,z\right)  =  0\,,
\end{eqnarray*}
where 
\begin{eqnarray}
R & = & \sqrt{x^{2}+y^{2}}\,,r=\sqrt{\left(R-1\right)^{2}+z^{2}}\,,\\
\bfe_{\phi} & = & [\frac{y}{R},-\frac{x}{R},0]\,,\bfe_{R}=[\frac{x}{R},\frac{y}{R},0]\,.
\end{eqnarray}
To analyze the particle flux in a static density distribution, we
considered two cases: case A and case B. In case A, we compared the neoclassical
particle flux
with the PS particle flux in the PS regime. The parameters are listed in Tab.~\ref{TabPPAB}.
\begin{table}
	\centering
	\begin{tabular}{|c|c|c|c|c|c|c|}
		\hline
		&$T_i$&$T_e$&$R_0$&$B_0$&$r_0$&$q$\\\hline
		Case A&$60\mathrm{eV}$&$60\mathrm{eV}$&$6\mathrm{m}$&$7\mathrm{T}$&$0.4\mathrm{m}$&$3.0$\\\hline
		Case B&$200\mathrm{eV}$&$200\mathrm{eV}$&$1.75\mathrm{m}$&$2.5\mathrm{T}$&$0.45\mathrm{m}$&$3.5$\\\hline
	\end{tabular}
	\caption{Plasma parameters used in case A and B.}
	\label{TabPPAB}
\end{table}
The particle source is located at $r=r_i=(r_\mathrm{left}+r_o)/2$,  while two sinks
are located at $r_\mathrm{left}$ and $r_o$ to maintain a static density distribution.
We calculated different background densities from 
$1.6\times10^{18}\mathrm{m}^{-3}$ to $1.6\times10^{21}\mathrm{m}^{-3}$ 
with different collisional frequencies, and both background ions and test particles are protons. 
We use three different pairs of $r_i$ and $r_o$, which are listed in 
Tab.~\ref{TBPSUIS}.
\begin{table}
	\centering
	\begin{tabular}{|c|c|c|c|}
		\hline
		&Group 1&Group 2&Group 3\\\hline
		Case A $r_i/\mathrm{cm}$&40&40&40\\\hline
		Case A $r_o/\mathrm{cm}$&50&46&42\\\hline
		Case B $r_i/\mathrm{cm}$&45&45&45\\\hline
		Case B $r_o/\mathrm{cm}$&53&48&46\\\hline
	\end{tabular}
	\caption{Particle sources and sinks used in simulations of Case A and B.}
	\label{TBPSUIS}
\end{table}
After a sufficient long simulation, we obtain the static density distribution 
and particle fluxes, shown in \FIG{FigFLUXVerifySTAT} (a) and (b), respectively. 
The neoclassical particle flux $\Gamma_\mathrm{NC}=\calD_\mathrm{NC}n'\left( r \right)$ with $n'\left( r \right)\approx (n_\mathrm{test}\left( r_o \right)-n_\mathrm{test}\left( r_i \right))/\left( r_o-r_i \right)$ is also plotted as reference.  When 
collisional frequency is high and the static density profile is steep, the 
particle flux violates the neoclassical theorem and saturates to $\GammaSP$, which is constent
with the new SP model. Furthermore, the 
classical diffusion will become dominant when its induced flux is comparable to $\GammaSP$.

In case B, we considered parameters that correspond to a real tokamak plasma, 
specifically the pedestal of 
EAST \cite{wu2007overview,gao2020experimental}.
The parameters used are also listed in Tab.~\ref{TabPPAB}.
The backgroud ion species is 
deterium and its density
varies from $1\EXP{18}$ to $1\EXP{20}$, test particles
are carbon ions, i.e., $\ce{ ^{12}_6C^{6+}}$. 
The source and sink setups are also listed in Tab.~\ref{TBPSUIS}. The collisional frequency 
for carbon ions is estimated as $\nu_\mathrm{C}=6^2\nu_\mathrm{D}$, where $\nu_\mathrm{D}$ is 
the collisional frequency of backgroud deteriums. The static density profile 
and particle fluxes are shown in \FIG{FigFLUXVerifySTAT} (c) and (d).  It is evident that when 
$n_0>1\EXP{19}\mathrm{m}^{-3}$, the particle flux of carbon ions is significantly lower 
than the prediction of the neoclassical theory and saturates to 
$\GammaSP$. In the
experiment reported in \REF{gao2020experimental} with $n_0\sim 2\EXP{19}\mathrm{m}^{-3}$, if 
$1/\kappa\sim3\mathrm{cm}$,
the SP particle flux is approximately one-third of the value predicted by neoclassical 
theory $\Gamma_\mathrm{NC}$.

In summary, our research has revealed that the neoclassical transport theory for toroidal plasmas fails to accurately predict particle transport when the collisional frequency $\nu$ is sufficiently high. This occurs when the quantity 
$G_\nu$, given by $G_\nu=2\nu R_0\kappa q^2 /\Omega_i$ 
approaches or becomes larger than $1$. We develop a new SP
model for investigating $G_\nu \sim 1$ and $G_\nu\gg 1$ regimes,
which concludes a new Bohm-like SP flux 
\begin{eqnarray}
	\GammaSP=\frac{4 n_0\left( r \right)T_i}{q_iB_0R_0\pi}\,.\label{EqnNEWTHEOREM}
\end{eqnarray}
It is proportional to the flux 
caused by VTF-drift, indicating a failure in the  
confinement by rotational transform. Importantly,
$\GammaSP$ is independent of the collisional frequency $\nu$, inverse density 
scale length $\kappa$, and the safety factor $q$, creating a type of transport 
barrier. Furthermore, it is proportional to $T_i/(B_0R_0)$ and exhibits a 
reduced Bohm-like scaling.

\begin{figure}[htp]
	\begin{center}
		\subfloat[Schema of PS transport for $G_\nu\ll1$.]{\includegraphics[width=0.49\linewidth]{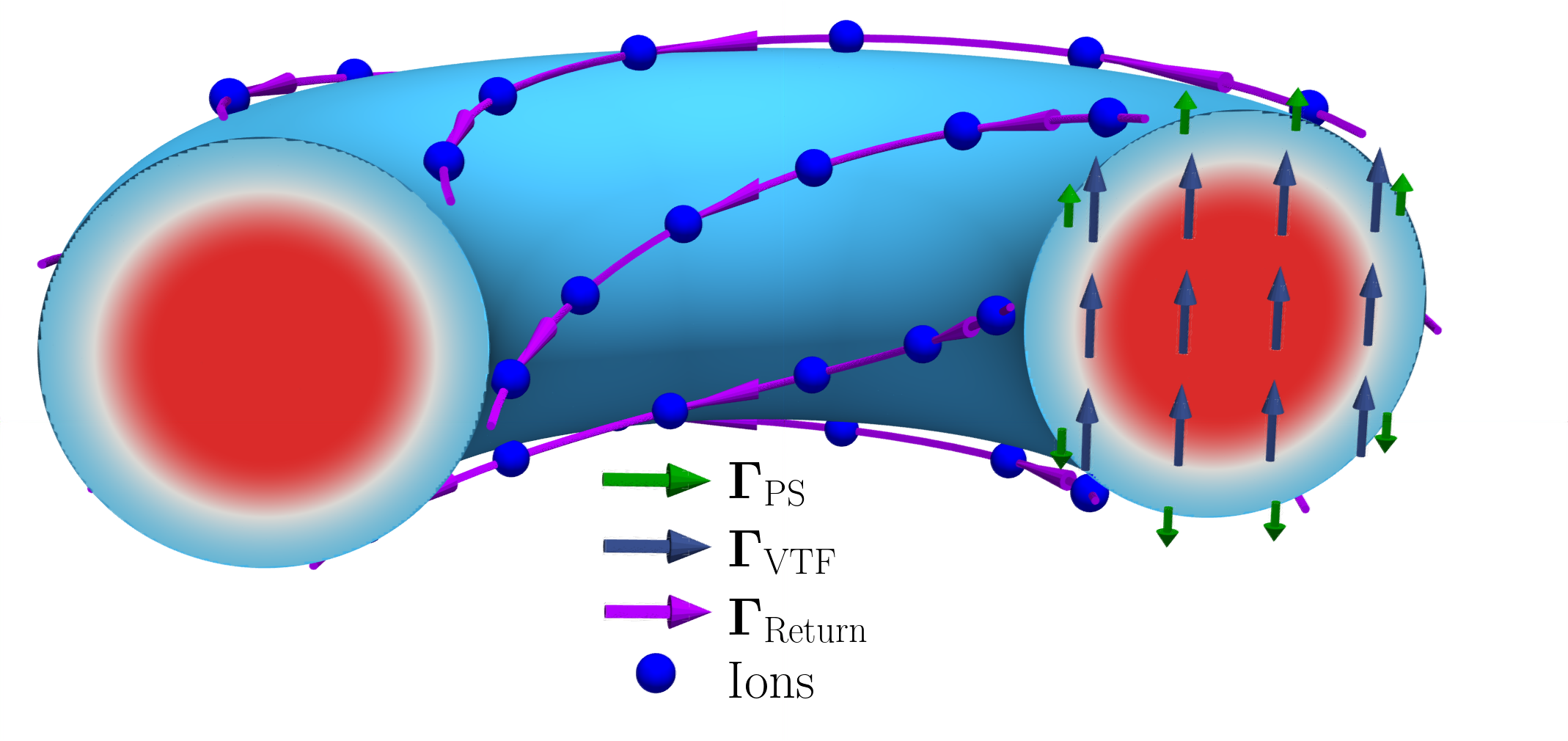}}
		\subfloat[Schema of SP transport for $1\ll G_\nu <q^2$.]{\includegraphics[width=0.49\linewidth]{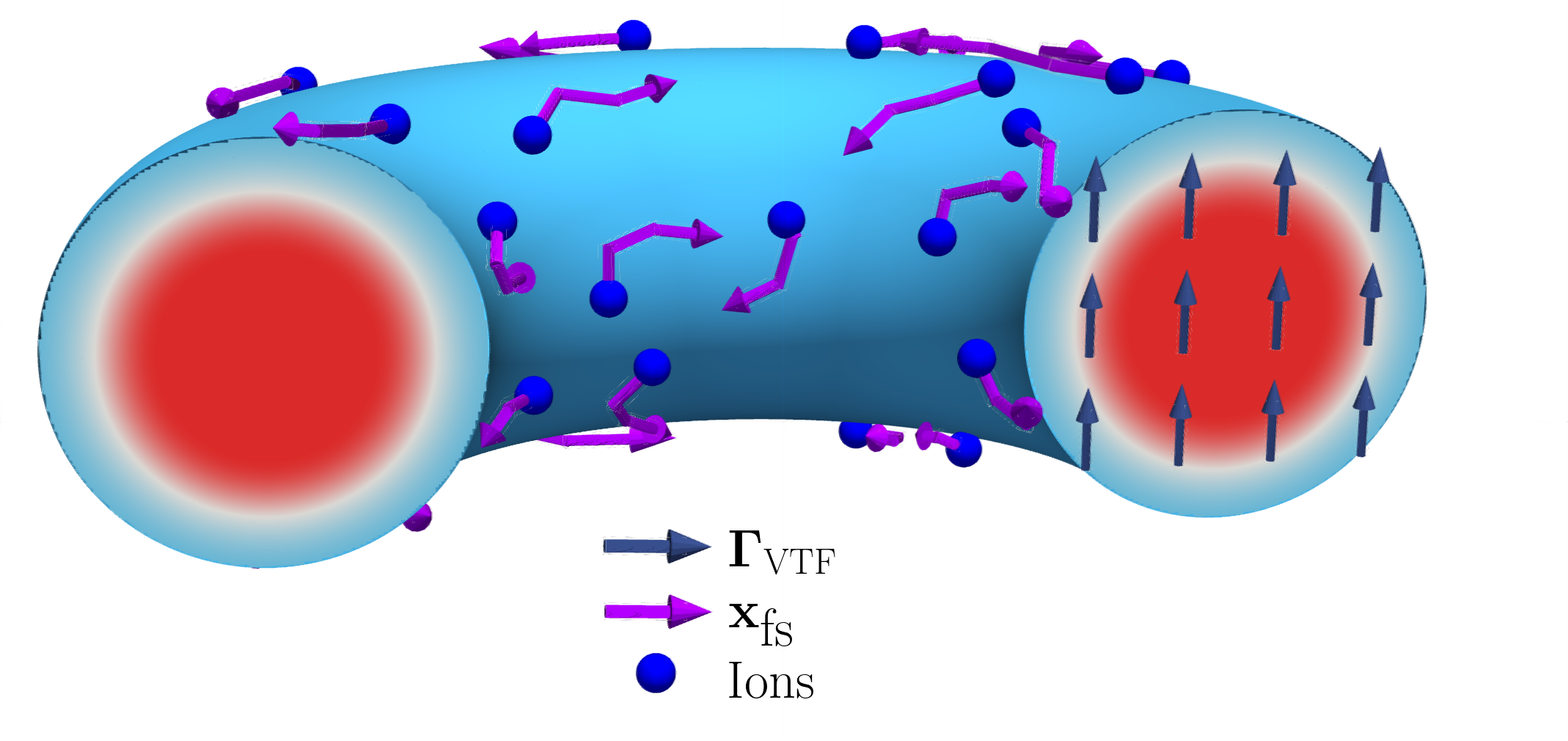}}
	\end{center}
	\caption{Principle schemas of High-$Z$ impurity particle flux 
	at the pedestal of tokamak plasma in the PS regime when $G_\nu\ll1$ (a) and for the 
	SP regime when $1\ll G_\nu<q^2$ (b). The direction 
	of VTF drift is upward. For the condition that conventional PS theory can be applied, 
	the relation $\nabla\cdot\left( \mathbf{\Gamma}_\mathrm{VTF}+\mathbf{\Gamma}_\mathrm{Return} \right)=0$ holds, 
	which means the final particle flux is $\mathbf{\Gamma}_\mathrm{PS}$. When $G_\nu\gg1$, the return flow will be disturbed by the collision (illustrated by $\bfx_\mathrm{fs}$), so the averaged SP particle flux is proportional to $\mathbf{\Gamma}_\mathrm{VTF}$.}
	\label{FigDiag}
\end{figure}

In real toroidal plasmas, this scenario could arise when considering the transport of 
high-$Z$ impurity ions at the tokamak pedestal, where the particle flux is dominated
by the VTF drift. Figure \ref{FigDiag} is a schema of SP transport, where 
the VTF drift direction is upward. According to the SP model, the particle flux is inward 
in the lower half of the plasma pedestal and outward in the upper half. If the impurity
source is uniformly distributed outside the plasma, the static distribution of impurity
ions should be asymmetric in the poloidal plane and the particle flux will saturate 
when the collisional frequency is sufficiently high. 
Such up-down asymmetric distribution (UDAD) of impurities has been observed in several 
experimental works
\cite{terry1977observation,rice1997x,reinke2013non}. 
In \REF{terry1977observation}, when
\begin{eqnarray}
	\frac{d_{\nabla B}}{\kappa}=\frac{3G_\nu}{4}\sim 1,
\end{eqnarray}
strong UDAD can be observed, which supports the new SP theory. The experimental parameters
in Refs.~\cite{rice1997x} and \cite{reinke2013non} and the occurrence of strong UDAD are shown in 
Tab.~\ref{TabPARAS}, which indicates that strong UDAD happens for $G_\nu>1$.
\begin{table}
	\centering
	\begin{tabular}{|c|c|c|c|c|c|c|c|c|c|c|}\hline
		Works&$T_i/\mathrm{eV}$&$R_0/\mathrm{m}$&$q$&$n_i/(10^{20}\mathrm{m}^{-3})$ & $\kappa\cdot\mathrm{m}$&$B_0/\mathrm{T}$&$Z$&$m_i/m_p$&$G_\nu$&UDAD\\\hline\hline
		Rice, et al. \cite{rice1997x} & 200 & 0.67&4.0 &0.5&33&5.3&17&40&15.9&Yes\\\hline
		Reinke, et al. \cite{reinke2013non}&1000$^*$&0.67&3.25$^*$&0.5&91&5.4&17&40&2.07&Yes\\\hline
		Reinke, et al. \cite{reinke2013non}&1000$^*$&0.67&2.0$^*$&0.6&75&5.4&17&40&0.78&No\\
\hline
	\end{tabular}
	\caption{Previous works reporting the UDAD of impurities in Tokamak plasmas. Some parameters marked by $^*$ are not provided directly and are derived from other parameters reported in corresponding works.}
	\label{TabPARAS}
\end{table}
To further validate the saturation property of the particle flux, we will analyze real 
experimental data and report related results in future works. 

This work is supported by the the National MC Energy R\&D Program
(2018YFE0304102), GuangHe Foundation (ghfund202202018672), Collaborative Innovation Program of Hefei Science Center, CAS, 
(2021HSC-CIP019),  the National Natural Science
Foundation of China (NSFC-11905220), National Magnetic Confinement Fusion Program of China (2019YFE03060000), Director Funding of Heifei Institutes of Physical Science from Chinese Academy of Sciences (Grant Nos. E25D0GZ5), National Natural Science Foundation of China (Grant Nos. E45D0GZ1) and Geo-Algorithmic Plasma Simulator (GAPS) Project.

\bibliographystyle{apsrev4-1}
\bibliography{bohm}

\end{document}